\documentclass[a4paper]{article}
\usepackage{color}
\usepackage{listings}
\usepackage{a4}
\usepackage[compact]{titlesec}

\usepackage{longtable}
\usepackage{amssymb}

\lstdefinestyle{defaultstyle}{}
\definecolor{lightgrey}{cmyk}{0.1,0.1,0.1,0}
\definecolor{grey}{cmyk}{0.5,0.5,0.5,0}

\def\va{{a}}
\def\vb{{b}}
\def\vc{{c}}
\def\vd{{d}}

\def\sa{{\alpha}}
\def\sb{{\beta}}

\def\ca{{\dot\alpha}}
\def\cb{{\dot\beta}}

\def\efb{{\tilde\mathcal B}}
\def\efc{{\tilde\mathcal C}}

\def\eaa{{\mathcal A}}

\begin{document}

\begin{center}
{\bf\Large Numerical tools to validate stationary points of $SO(8)$-gauged $\mathcal{N}=8$ $D=4$ supergravity}

\bigbreak

{\bf Thomas Fischbacher\\}
\smallbreak
{\em University of Southampton\\
  School of Engineering Sciences\\
  Highfield Campus\\
  University Road, SO17 1BJ Southampton, United Kingdom\\}
{\small {\tt t.fischbacher@soton.ac.uk}}

\end{center}

\begin{abstract}
  \noindent Until recently, the preferred strategy to identify
  stationary points in the scalar potential of $SO(8)$-gauged
  $\mathcal{N}=8$ supergravity in $D=4$ has been to consider
  truncations of the potential to sub-manifolds of $E_{7(+7)}/SU(8)$
  that are invariant under some postulated residual gauge group
  $G\subset SO(8)$. As powerful alternative strategies have been shown
  to exist that allow one to go far beyond what this method can
  achieve -- and in particular have produced numerous solutions that
  break the $SO(8)$ gauge group to no continuous residual symmetry --
  independent verification of results becomes a problem due to both
  the complexity of the scalar potential and the large number of new
  solutions.  This article introduces a conceptually simple
  self-contained piece of computer code that allows independent
  numerical validation of claims on the locations of newly discovered
  stationary points.
\end{abstract}

\section{Introduction}

\noindent Maximally extended ($\mathcal{N}=8$) supergravity in four
dimensions recently has seen a resurgence of interest, in particular
due to speculations about its potential perturbative
finiteness~\cite{Bern:2009kd,Bianchi:2009wj,Dixon:2010gz}, as well as
the ${\rm AdS_4/CFT_3}$ correspondence in the context of ABJM/BLG
theory~\cite{Benna:2008zy} and also early ideas about an
AdS/CMT correspondence~\cite{Hartnoll:2008vx,Bobev:2010ib}.


As is well known, $\mathcal{N}=8$ supergravity allows the promotion of
its 28~vector fields to nonabelian gauge fields. This was first
demonstrated for the gauge group $SO(8)$ in~\cite{de Wit:1981eq}. As
has been shown in~\cite{Hull:1984qz, Hull:2002cv}, there are other,
more exotic, options as well; in particular, it is possible here to
introduce \emph{noncompact} gauge groups such as $SO(p,8-p)$.

In order to maintain supersymmetry, such a deformation of the model
requires the introduction of a potential on the 70-dimensional
manifold of scalars. This potential is known to have a number of
nontrivial stationary points, some of which correspond to stable
vacuum states with broken symmetry and usually some residual
supersymmetry. Unfortunately, the analysis of these potentials is
fairly involved, as they are most naturally regarded as functions on
the coset manifold $E_{7+7)}/SU(8)$, which is infeasible to
parametrize analytically by introducing coordinates, as is readily
seen by simple order-of-magnitude guesstimations on the number of
expected combinations of trigonometric and hyperbolic function factors
in a generalized Euler angle parametrization.

For a long time, the favored approach to nevertheless extract some
information about stationary points was to consider restrictions of
the potential to sub-mainfolds of $E_{7+7)}/SU(8)$ that are invariant
under some subgroup $G$ of the gauge group, the rationale being that
the stationarity condition on the restricted manifold readily is
lifted to the full manifold, as the first order term in a perturbative
expansion of the potential must be a $G$-invariant scalar. While
further technical subtleties such as coordinate singularities might
complicate the analysis, this has been shown to be a powerful
technique that allowed the determination of all $SU(3)$-invariant
stationary points~\cite{Warner:1983vz}, as well as a further
stationary point with residual symmetry~$SO(3)\times SO(3)$,
cf.~\cite{Warner:1983du}. No further solutions have been found until a
new strategy has been proposed in~\cite{Fischbacher:2009cj} that
increased the number of known nontrivial solutions from six to twenty.

While this method, which is based on numerical techniques that were
discovered at about the same time~\cite{Speelpenning1980} as the scalar
potentials of gauged maximal supergravity, allows a much deeper
analysis and is expected to give rise to numerous other solutions -- as
it did~\cite{Fischbacher:2008zu} for $SO(8)\times SO(8)$-gauged Chern-Simons
$\mathcal{N}=16$ supergravity in~$D=3$~\cite{Nicolai:2000sc} -- it
raises two new issues: First, this is `only' a numerical
approach that provides evidence for new stationary points, but no
stringent proof of their existence. And second, due to the large
number of solutions and the analytic complexity of these potentials,
checking these claims unfortunately amounts to a substantial
task.

The first issue can be addressed in a semi-automatic way by employing
another modern and fairly revolutionary algorithm -- the the PSLQ
`lattice shift' algorithm to find integer relations between numbers
given with high-precision~\cite{PSLQ}. The basic principle is
demonstrated in section~7 of~\cite{Fischbacher:2010ec} and also
briefly explained in section~\ref{sec:outlook}. As obtaining a very
large number of valid digits is computationally expensive, the cut-off
parameters for the PSLQ algorithm will generally require some manual
tweaking, so striving for full automatization of this step, while
possible, would probably not be a worthwhile objective. The
post-processing of analytic conjectures obtained via the PSLQ
algorithm, which amounts to establishing their analytic validity,
should be fairly straightforward to automatize, however. The challenge
is to analytically exponentiate exact expressions for Lie algebra
elements using symbolic algebra (probably somewhat tailored to the
algebraic task) and then symbolically verifying the stationarity
condition.

The second issue is addressed by this article, which provides computer
code to numerically validate claims about new stationary points of
$SO(8)$-gauged $\mathcal{N}=8$ supergravity with little effort. As it
is expected that many more solutions and maybe also other, even more
powerful methods to analyze such scalar potentials could be discovered
in the future, having a self-contained dedicated tool for such a task
-- as is presented in this article -- seems to make sense.

\section{The Code}

\noindent Given the main objective of validation of stationary points
in the scalar potential of $SO(8)$-gauged $\mathcal{N}=8$
supergravity, computer code that calculates this potential should be
clean, compact, self-contained, and easily verifiable. This is best
achieved by an independent re-implementation of numerical code that
emphasizes simplicity and verifiability (i.e. a close match between
mathematical formulae and code) over algorithmic tricks to achieve
high performance and also is independent of the codebase that was used
to find the new solutions presented
in~\cite{Fischbacher:2009cj}\footnote{That codebase is at the time of
  this writing a complex multi-language mix of LISP and Python modules
  that mostly deal with problems other than validating results and are
  not even available on all computing platforms (in particular
  Microsoft Windows). It will be provided after major clean-up work to
  make it easy to install also on non-Unix platforms.}.

\subsection{Design Choices}

\noindent The problem of numerically computing the supergravity scalar
potential -- while involving some intricate definitions -- only
requires arithmetics, a small bit of combinatorics, and as the only
tricky step, exponentiation of a complex matrix. Therefore, pretty
much any programming language could be used as an implementation
basis. While problems of such a nature that do not require symbolic
transformations often are addressed with MATLAB~\cite{MATLAB}, or its
free alternative GNU Octave~\cite{Octave}, Python~\cite{Python} has
been chosen for this work, for the following reasons:

\begin{itemize}

\item Python allows a transcription of the required combinatorics to
  computer code with considerably less overhead than many other
  languages (but not as effectively as Lisp or Scheme).

\item Python is free (in contradistinction to commercial computer algebra
  packages) and readily available across a broad range of hardware and
  operating system platforms.

\item Python syntax is very simple and conceptually sufficiently
  similar to other programming languages popular with casual
  programmers that it can be reasonably expected to be comprehensible
  to a broader audience than many other languages.

\item Python provides a command prompt that allows interactive
  evaluation of code.

\item Python itself has become somewhat popular for many scientific
  problems.

\end{itemize}

A major architectural drawback of the code presented here is that, as
it deliberately is kept algorithmically simple, it neither uses
sophisticated combinatorics nor efficient black-box approaches (such
as relational database algorithms for sparse tensors, as the
LambdaTensor~\cite{Fischbacher:2002fr} package does) to keep the
computational effort to a minimum. However, for the purpose of
validating results, speed is not an important concern.

\subsection{Installation and Usage}

\noindent The codebase utilizes the Python modules \texttt{numpy} for
(non-sparse) tensor numerics as well as \texttt{scipy} (Scientific
Python) for matrix inversion and exponentiation, and hence requires
both Python as well as these extension modules to be installed.  Most
free Unix distributions are strongly modularized and provide the
corresponding packages; for other platforms such as Microsoft Windows,
convenient pre-bundled Python packages that contain these modules are
available. Installation of Interactive Python (\texttt{ipython}) also
is highly recommended.

Once Python is installed and the source that accompanies this article
is downloaded from \texttt{http://arxiv.org/e-print/1007.0600} and
unpacked, the functions described below can be made available by
importing the corresponding module into Python. The most important
constant certainly is the dictionary \texttt{vacua\_v70} from
\texttt{e7\_vacua.py} -- which contains all the known stationary
points. The dictionary keys are strings of the form `\texttt{\#0}',
`\texttt{\#1}', `\texttt{\#2}', etc. that correspond to the numbers
used in the tables in~\cite{Fischbacher:2009cj}. All Python functions
come with interactive online help. A transcript of an interactive
Python session is given in figure~\ref{fig:pyinteractive}.

\begin{figure}
{\small
\protect\begin{lstlisting}
$ ipython
In [1]: from e7_analysis import *
# (This takes a while, as definitions have to be computed)

In [2]: from e7_vacua import *

In [3]: help(e7_A1_eigenvalues)
# This brings up online help for this function

In [4]: e7_A1_eigenvalues(vacua_v70["#3"])
Out[4]: 
(array([ 1.33758921-0.09352799j,  1.33758921-0.09352799j,
        1.09478134+0.00705399j,  1.33758921-0.09352799j,
        1.33758921-0.09352799j,  1.33758921-0.09352799j,
        1.33758921-0.09352799j,  1.33758921-0.09352799j]),
 [1.4999987205717096,
  1.4999987205717116,
  1.0000000000469873,
  1.4999987205717116,
  1.4999987205717089,
  1.4999987205717109,
  1.4999987205717116,
  1.4999987205717105])

In [5]: e7_test_stationarity_Q(vacua_v70["SU(4)"])
Out[5]: 3.0086481280021352e-15
# This shows that the Q-tensor stationarity condition is
# satisfied to high numerical accuracy for the known
# stationary point with SU(4) symmetry.
\end{lstlisting}}
\caption{\label{fig:pyinteractive}An interactive \texttt{ipython} session 
with the validation code.}
\end{figure}

For some investigations, scripting is preferable to interactive
usage. This is most easily done by adding new Python files to the code
directory; alternatively, one can extend the \texttt{PYTHONPATH}
environment variable to tell Python where to look for modules and put
additional code into an arbitrary directory. The file
\texttt{e7\_example.py} which also is listed in
figure~\ref{fig:pyexample} shows scripting use. It is executed as
`\texttt{python e7\_example.py}'.

\begin{figure}
\begin{minipage}{1.0\textwidth}
{\small
\begin{lstlisting}
from e7_vacua import *
from e7_analysis import *

for n in range(21):
    key="#"+str(n)
    v70=vacua_v70[key]
    (gens,svd)=e7_residual_gauge_group_generators(v70)

    print "Solution:", key,\
          "P=",e7_potential_from_v70(v70),\
          "Q=",e7_test_stationarity_Q(v70),\
          "dim(res. GG)=",len(gens)
\end{lstlisting}
}
\end{minipage}
\begin{minipage}{1.0\textwidth}
{\scriptsize
\begin{lstlisting}
Solution: #0 P= -6.0 Q= 0.0 dim(res. GG)= 28
Solution: #1 P= -6.68740304976 Q= 2.12995452535e-15 dim(res. GG)= 21
Solution: #2 P= -6.98771242969 Q= 5.07925733935e-15 dim(res. GG)= 21
Solution: #3 P= -7.19157562605 Q= 6.61364975244e-06 dim(res. GG)= 14
Solution: #4 P= -7.79422863406 Q= 3.98414774796e-15 dim(res. GG)= 9
Solution: #5 P= -8.0 Q= 3.008648128e-15 dim(res. GG)= 15
Solution: #6 P= -14.0 Q= 2.98299102544e-15 dim(res. GG)= 6
Solution: #7 P= -9.98708340034 Q= 1.69023258423e-06 dim(res. GG)= 1
Solution: #8 P= -10.434712595 Q= 4.19318631751e-08 dim(res. GG)= 0
Solution: #9 P= -10.6747542478 Q= 7.60840475615e-05 dim(res. GG)= 2
Solution: #10 P= -11.6568542495 Q= 3.12755317279e-07 dim(res. GG)= 2
Solution: #11 P= -11.9999999998 Q= 8.08047489882e-07 dim(res. GG)= 2
Solution: #12 P= -13.6236525917 Q= 1.2589806911e-06 dim(res. GG)= 1
Solution: #13 P= -13.6761142184 Q= 5.80889427017e-07 dim(res. GG)= 0
Solution: #14 P= -14.97038467 Q= 1.4218350389e-07 dim(res. GG)= 1
Solution: #15 P= -16.414456312 Q= 9.21761357658e-06 dim(res. GG)= 0
Solution: #16 P= -17.8764434504 Q= 0.0010604463244 dim(res. GG)= 0
Solution: #17 P= -18.0526932154 Q= 1.30960846002e-06 dim(res. GG)= 0
Solution: #18 P= -21.2659762563 Q= 6.9908080844e-07 dim(res. GG)= 0
Solution: #19 P= -21.4084980043 Q= 0.000232765771 dim(res. GG)= 0
Solution: #20 P= -25.1493689791 Q= 5.05511356071e-07 dim(res. GG)= 0
\end{lstlisting}
}
\end{minipage}
\caption{Example code demonstrating scripting use of the
validation code and its output.\label{fig:pyexample}}
\end{figure}

\subsection{Modules and Functions}

\noindent The code itself is modularized and consists of these components:

\subsubsection{The \texttt{tensor\_io.py} module}

\noindent This component defines functions that read and write
numerical (higher-rank) tensors in a well-defined simple sparse
textual data format that is also human-readable.  other codebases. It
is generally expected that programs which manipulate tensors might
want to save these to files, both for persistent data storage and data
exchange. The full format definition, which may be adopted by other
programs, is given in the Python documentation of this module. It
provides the functions \texttt{tensor\_print()},
\texttt{tensor\_read()}, and \texttt{tensor\_write()}.

\subsubsection{The \texttt{e7\_combinatorics.py} module}

\noindent This component provides a few combinatorics-related
definitions that are of use to other components. In particular, it
defines two-index and four-index ranges running from $(0,0)$ to
$(7,7)$ and $(0,0,0,0)$ to $(7,7,7,7)$, respectively. Also, it
introduces iteration over 3-permutations and 4-permutations. Here, it
has to be kept in mind that index counting starts at 0, not
1. Matching published data to numerical results hence requires
shifting all index ranges found in the literature by one.

\subsubsection{The \texttt{e7\_definitions.py} module}

\noindent This component provides definitions related to the $\rm
spin(8)$ algebra that are then used to define explicit forms for the
133~complex $56\times 56$ $E_{7(+7)}$ generator matrices of the
fundamental representation. The most important definitions are these
non-sparse \texttt{numpy} higher-rank arrays:

\begin{itemize}
\item \texttt{T\_e7} -- the $133\times56\times56$ tensor
  $T^{(E7)}{}_{\eaa\efb}{}^{\efc}$ from formula $(A.12)$ in
  \cite{Fischbacher:2009cj}.

\item \texttt{so8\_gamma\_a\_aS\_aC} -- the $8\times8\times8$ tensor
  $\gamma^\va{}_{\sa\ca}$ containing ${\rm spin(8)}$ `gamma
  matrices'. The order of indices is vector, spinor, co-spinor, as
  indicated by the name.

\item \texttt{so8\_gamma\_ab\_aS\_bS},
  \texttt{so8\_gamma\_ab\_aC\_bC},\\
  \texttt{so8\_gamma\_abcd\_aS\_bS},
  \texttt{so8\_gamma\_abcd\_aS\_bS} -- likewise, these (non-sparse)
  $8\times8\times8\times8$ and $8\times8\times8\times8\times8\times8$
  arrays give the tensors
  $\gamma^{\va\vb}{}_{\sa\sb}$, $\gamma^{\va\vb}{}_{\ca\cb}$,
  $\gamma^{\va\vb\vc\vd}{}_{\sa\sb}$, and
  $\gamma^{\va\vb\vc\vd}{}_{\ca\cb}$.
\end{itemize}

Furthermore, there are functions to map numerical vectors of length~70
that contain ${\bf 35}_{\rm s}+{\bf 35}_{\rm c}$ coefficients of the
`boost' generators of $E_{7(+7)}$ to the language of self-dual and
anti-self-dual four-forms and back. Internally, four-form coefficients
are represented as Python dictionaries (hash tables). The
\texttt{format\_abcd\_to\_string()} and
\texttt{format\_abcd\_to\_latex()} functions can be used to bring
these into human-readable form.

The function \texttt{abcd\_from\_v70()} maps a $70$-vector to a
four-form Python dictionary. Its inverse is given
by~\texttt{v70\_from\_abcd()}. Normalization conventions are such that
if a $70$-vector $v$ has a single non-zero entry, then the non-zero
entries (192 of them) in the corresponding four-form dictionary are
all $\pm1$.

\subsubsection{The \texttt{e7\_potential.py} module}

\noindent This component contains the definitions needed to calculate
the supergravity scalar potential. The most important function is
\texttt{e7\_potential\_from\_v70()}, which computes and returns the
potential to numerical accuracy. As this function can also return the
most important intermediate quantities used in the calculation, such
as the tensors $A_1$ and $A_2$, when called with an additional
dictionary parameter \texttt{info} to store data in, the other
auxiliary functions defined here -- such as \texttt{e7\_A12()} --
are usually not needed.

Based on the potential function, the function
\texttt{e7\_test\_gradient()} uses finite differences to numerically
calculate an approximation to the potential's gradient as well as its
length at a given point. This provides a direct test for stationarity
and hence allows the numerical validation of solutions.

\subsubsection{The \texttt{e7\_vacua.py} module}

\noindent This module provides just a single constant, the dictionary
\texttt{vacua\_v70} that contains 70-vectors for all the known
stationary points in the scalar potential. Some solutions are
associated with named keys, such as `\texttt{G2}', and all of them can
be addressed with keys of the form `\texttt{\#17}', corresponding to
the list given in~\cite{Fischbacher:2009cj}.

This module also provides a function \texttt{e7\_verify\_vacua()} that
re-does the automatic gradient check for all known stationary points.
This function uses the function \texttt{e7\_test\_gradient} from the
\texttt{e7\_potential.py} module.

\subsubsection{The \texttt{e7\_analysis.py} module}

\noindent This module provides additional functions to analyze
the properties of stationary points. The most important ones are:

\begin{itemize}

\item \texttt{e7\_test\_stationarity\_Q()} -- to verify whether a
  point satisfies the $Q_{abcd}+\frac{1}{24}\eta\epsilon_{abcdefgh}Q^{efgh}=0$
  stationarity condition. This provides an independent test in
  addition to the direct numerical evaluation of the gradient provided by
  \texttt{e7\_test\_gradient()}.

\item \texttt{e7\_residual\_gauge\_group\_generators()} -- to
  automatically determine a basis of the subgroup of
  $\rm spin(8)$ that is left unbroken by the point in question.

\item \texttt{e7\_susy\_A2()} -- to numerically determine the number
  of unbroken supersymmetries via the $\epsilon^i  A_{2\;i}{}^{jk\ell}=0$ condition.

\item \texttt{e7\_A1\_eigenvalues()} -- to numerically determine the
  eigenvalues of the $A_1$ tensor. In addition to these eigenvalues
  $E$, the values $-6|E|^2/P$ are returned as well. This gives a
  second test for unbroken supersymmetry.

\item \texttt{e7\_scalar\_massmatrix()} -- to numerically calculate
  the re-scaled scalar mass matrix given by formula $(2.25)$ in
  \cite{deWit:1983gs}:
  \begin{eqnarray*}
  96e^{-1}g^{-2}L_M &=&
  -\left(\frac{2}{3}\mathcal{P}(\mathcal{V})+\frac{13}{72}A_2{}^m{}_{npq}A_2{}_m{}^{npq}\right)\Sigma_{ijkl}\Sigma^{ijkl}\\
  &&-\left(6A_2{}_k{}^{mni}
    A_2{}^{j}{}_{mn\ell}-\frac{3}{2}A_2{}_n{}^{mij}A_2{}^n{}_{mk\ell}\right)\Sigma_{ijpq}\Sigma^{k\ell pq}\\
  &&+\frac{2}{3}A_2{}^i{}_{mnp}A_2{}_q{}^{jk\ell}\Sigma^{mnpq}\Sigma_{ijk\ell}.
  \end{eqnarray*}
  One normally will want to re-scale this matrix by multiplying with
  $-6/\mathcal{P}$ ($\mathcal{P}$ being the value of the potential) in
  order to obtain masses in AdS-units. After such re-scaling, the
  Breitenlohner-Freedman stability bound~\cite{Breitenlohner:1982jf}
  is $-(d-1)^2/4=-9/4=-2.25$.

\item \texttt{e7\_A12\_masses2()} -- to calculate the fermion
  masses-squared according to formula $(2.14)$ in \cite{deWit:1983gs}.

\end{itemize}

\section{Outlook}
\label{sec:outlook}
\noindent Deeper numerical investigations than those presented
in~\cite{Fischbacher:2009cj} suggest that the number of yet
unpublished stationary points in the scalar potential studied here as
well as related ones may be fairly high. A number of these solutions
may easily turn out to be sufficiently interesting to warrant a deeper
investigation. One might think that the idea of using numerical
techniques to find approximate solutions should be of very limited use
in particular for those solutions that break the gauge group (almost)
completely: Studying a stationary point numerically in order to get
ideas that may guide a fully analytic approach seems to be a fairly
unrewarding exercise unless it has a sufficient degree of residual
symmetry. However, this is not necessarily the case. First of all, it
is possible to eliminate almost all of the arbitrariness in the raw
numbers that is associated with the freedom to apply a ${\rm
  spin(8)}$-rotation to a given solution, using the techniques that
were employed in~\cite{Fischbacher:2008zu,Fischbacher:2009cj} to
minimize the number of non-zero coefficient vector entries. There is
no fundamental obstacle to modifying readily existing techniques to
use very high precision arithmetics (with hundreds of digits) rather
than machine-precision 16-digit arithmetics in order to determine
parameters very precisely. This then allows one to utilize powerful
integer relation algorithms such as in particular
PSLQ~\cite{PSLQ} to systematically `guess' even very complicated
analytic expressions. To demonstrate this, let us specifically consider a
number such as the coordinate $z_{(2)}$ from table A.1
in~\cite{Bobev:2010ib}: this is given as:

\begin{equation}
z_{(2)}=\frac{1}{4}\left(1-\frac{i}{3^{1/4}}\sqrt{2+\sqrt{3}}\right)\left(3+\sqrt{3}-3^{1/4}\sqrt{10}\right)
\end{equation}

To 300 digits, the imaginary part of $z_{(2)}$ is:

\begin{equation}
\begin{array}{lcl}
{\rm Im}\,z_{(2)}&\approx&
  -0.209269477042954393209112597384263059035631424\\
&&839824728046049646351961493949022694957040180268\\
&&460854129215451425779247977077536023519697290138\\
&&356018655716825868238074107476584372107488197263\\
&&113362737435710197764338927716172598148291052464\\
&&018329383941096143123970705213689657726314686822\\
&&409336284550127068
\end{array}
\end{equation}

If we truncate this number to 200 digits and use the PSLQ algorithm to
then find a vector of integer coefficients for integer powers of this
number so that the sum is closer to zero than $10^{-170}$, the PSLQ
algorithm produces the following polynomial coefficients:

\begin{equation}
[-64, 0, 704, 0, -240, 0, 32, 0, -1]
\end{equation}

One could say that this analytic expression allows us to make a
testable `prediction' for the next 100 digits. In fact, using Newton's
algorithm to find the zero of interest of this polynomial reproduces
${\rm Im}\,z_{(2)}$ to 100 digits beyond the previous 200. This is
demonstrated in detail the example code file
 \texttt{e7\_example\_mpmath.py}. In order to run this piece of code,
 the `\texttt{mpmath}' Python extension~\cite{mpmath} has to be installed.

It seems hence very likely that even for new stationary points that
can be expected to require even more complicated analytic
expressions, obtaining -- and most likely even rigorously proving --
analytic results can be fully mechanized.

The codebase that has been presented in this work (a) correctly
reproduces the known properties of the `classical' non-trivial
stationary points, (b) should be simple enough to be readily validated
against the formulae given in~\cite{Fischbacher:2009cj}, and (c)
should in the future help to considerably reduce the effort required
to check claims about the properties of new stationary points.

\paragraph{Acknowledgments}

It is a pleasure to thank N. Warner and K. Pilch for help with the
\texttt{e7\_scalar\_massmatrix} function.

\end{document}